\documentclass[9pt,twocolumn]{article}
\usepackage{amsfonts}
\usepackage{amsmath}
\usepackage{graphicx}
\usepackage[section]{placeins}
\usepackage{dcolumn}
\usepackage{hyperref}
\usepackage[T1]{fontenc}
\usepackage[utf8]{inputenc}
\usepackage{authblk}

\title{Automatic sorting of point pattern sets using Minkowski Functionals}
\author[1]{Joshua Parker}
\author[2]{Eilon Sherman}
\author[3]{Matthias van de Raa}
\author[3]{Devaraj van der Meer}
\author[4]{Lawrence E. Samelson}
\author[1]{Wolfgang Losert}
\affil[1]{Department of Physics, University of Maryland, USA}
\affil[2]{Racah Institute of Physics, The Hebrew University of Jerusalem, Israel}
\affil[3]{Institute for Nanotechnology, University of Twente, The Netherlands} 
\affil[4]{Center for Cancer Research, The National Institutes of Health, USA}

\begin{document}
\maketitle


\begin{abstract}
Point pattern sets arise in many different areas of physical, biological, and applied research, representing many random realizations of underlying pattern formation mechanisms. These pattern sets can be heterogeneous with respect to underlying spatial processes, which may not be visually distiguishable. This heterogeneity can be elucidated by looking at statistical measures of the patterns sets and using these measures to divide the pattern set into distinct groups representing like spatial processes. We introduce here a numerical procedure for sorting point pattern sets into spatially homogenous groups using Functional Principal Component Analysis (FPCA) applied to the approximated Minkowski functionals of each pattern. We demonstrate that this procedure correctly sorts pattern sets into similar groups both when the patterns are drawn from similar processes and when the 2nd-order characteristics of the pattern are identical. We highlight this routine for distinguishing the molecular patterning of fluorescently labeled cell membrane proteins, a subject of much interest in studies investigating complex spatial signaling patterns involved in the human immune response. 
\end{abstract}

\section{Introduction}
\subsection{Motivation: Why study point patterns?}
Spatial points patterns naturally arise in many areas of research in both the physical and life sciences, including ecology \cite{ecology, wiegand}, crime statistics \cite{crime}, epidemeology \cite{epid}, economics \cite{econ}, seismology \cite{seis}, material science \cite{material}, and astronomy \cite{astro}. Whether the points represent molecules, trees, cell phone users, or entire galaxies, the spatial distributions of point patterns belie the underlying stochastic processes that govern the pattern's formation.

\begin{figure}[ht]
\includegraphics[width=0.48\textwidth]{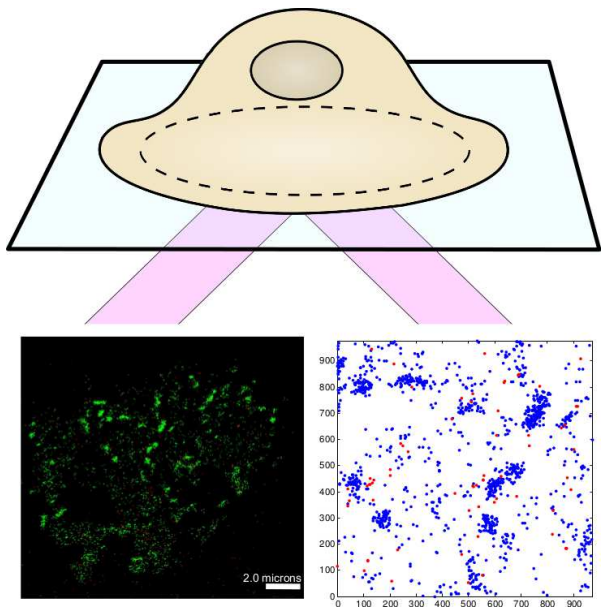}
\caption{\label{experiment} Using photoactived localization microscopy, the fluorescenctly labeled proteins are localized by fitting a point spread function to the stochastically photoactived molecules; the final pattern represents thousands of fluoresecent images}
\end{figure}

A new area of point pattern analysis involves studying the molecular patterning of proteins on the surfaces of cells. Due to photo-activated localization microscopy (PALM) \cite{PALM}, a new super-resolution microscopy technique, cell biologists are now able to measure the spatial distribution of fluorescently-tagged membrane proteins and determine the response of the molecules to different stimuli (see figure \ref{experiment}). By fixing the cells on a slide and exposing them to laser light, researchers can activate molecules one by one in multiple cells, locating the center to within 20 nm. This new technique has resulted in a wealth of new point pattern data representing different molecules and surface treatments, and quantitative analysis of these patterns can contribute much to understanding protein-protein and protein-membrane interactions \cite{coloc, Eilon}.
From a theoretical standpoint, each pattern is a pure realization of an underlying spatial process and can be used to characterize that process. From a practical perspective, however, it takes many experimental realizations with finite systems to discern the underlying structure. Furthermore, if the point interactions are complex or the patterns are formed in complicated environments (such as the membrane of a cell), the amount of data needed to confidently quantify a process becomes large and cumbersome to analyze. This gives rise to the need to be able to confidentely divide large sets of patterns, sorting the patterns into smaller, homogenous groups that can be analyzed further. In addition to simplifying analysis, this type of sorting can also provide researchers with quick information about the homogeneity of a process and the experimental parameters affect this homogeneity. 

\FloatBarrier
\subsection{Current Methods for Sorting Patterns}
The standard method of sorting pattern sets is as follows: For each pattern, one calculates a list of numerical summary characteristics (e.g. Index of Dispersion, Clark-Evans index). These can be regarded as the ''coordinates" of a pattern, to which distance-based clustering algorithms can be applied \cite{Stat}. This approach presents the researcher with the task of deciding which characteristics to use, how to compare them (normalizing, z-cores, etc) as well as how many: too few may result in missing information, too many could result in redundancy. This adds nuance to the sorting, limiting the statistical conclusions that can be drawn, and making trustable automation of the sorting procedure for large pattern sets difficult to accomplish.

A more robust sorting technique has been developed, where patterns are sorted using Functional Principal Component Analysis (FPCA) on smoothed 2nd-order functionals of the patterns. This routine treats the point set as a set of functions, $\{a_{i}(t)\}$, such as the pairwise correlation function, $g(r)$ \cite{FPCA}. The coordinates of each pattern are then calculated by finding the eigenfunctions and corresponding eigenvalues of the equation:
\begin{equation}
\int v(s,t)w_i(s)ds = \lambda_i w_i(t).
\end{equation}
Here $v(s,t)$ is the variance-covariance function of the set of functionals $a(r)$, defined as
\begin{equation}
v(s,t) = (N-1)^{-1}\displaystyle\sum\limits_{i=0}^N (a_{i}-\bar{a}_i(t))(a_{i}(s)-\bar{a}_{i}(s)).
\end{equation}
The ''score" of the $i$-th pattern on the $j$-th principal component is then $\int a_i(t)w_j(t)dt$ (see \cite{ramsay1,ramsay2}). Like standard PCA, the eigenvalues form a positive decreasing set whose truncated sum represent the total variance encapsulated in the included principal components. For automation, one can simply set a threshold for the amount of variance to be included, which in turn prescribes the number of coordinates to be used. This feature removes the arbitrariness of sorting patterns via the standard method, making FPCA an easily automatable way of quantifying the difference between patterns. 

However, spatial processes can create patterns with more structure than 2nd-order functionals can measure. The Neyman-Scott process (NS), introduced to study galaxy clustering, involves randomly distributed parent points generating clusters of varying size. The complexity of the parent/daughter interaction gives rise to families of NS processes with the same pairwise correlation function \cite{diggle,stoyan,Stat}, despite underlying spatial differences in the patterns.

Baddeley and Silverman also introduced a cell process which is built by partitioning a domain into cells of equal size which are then filled with a varying number of uniformly distributed points. Though the process is rather regular, they showed analytically that their process was indistinguishable from a Poisson process when only considering 2nd-order functionals of the pattern \cite{baddelley}, meaning that higher-order functionals must be used to resolve this ambiguity.

\subsection{From points to discs}
In this paper, we apply the proximity measure of FPCA to the approximated Minkowski functionals of point patterns \cite{Mink}. These functionals are calculated by centering a disc on each point and analyzing the topology of this secondary pattern of overlapping discs as a function of the radius. Since the overlap can be very complex, involving all possible combinations of individual points, these functionals depend on all orders of interaction simultaneously. This makes them a more complete ''fingerprint" for pattern comparison \cite{Mink,cosmo}. These functionals have enjoyed marked success in astrophysics \cite{aa}, soft matter \cite{alum}, and fluid turbulence \cite{detlef}. 

For completeness, we first explain the Minkowski functionals and how they are applied to point pattern analysis. We then demonstrate the sorting procedure by clustering sets of patterns of both synthetically generated data as well as biological data representing the spatial distributions of membrane proteins. Using both agglomerative and divisive clustering algorithms, we show that this procedure outperforms FPCA clustering with 2nd-order functionals, and in general we demonstrate it to be a viable method for automatically sorting point pattern sets.
\section{Outlining the Procedure}
\subsection{Minkowski functional analysis of point patterns}
\begin{figure}
\includegraphics[width=0.48\textwidth]{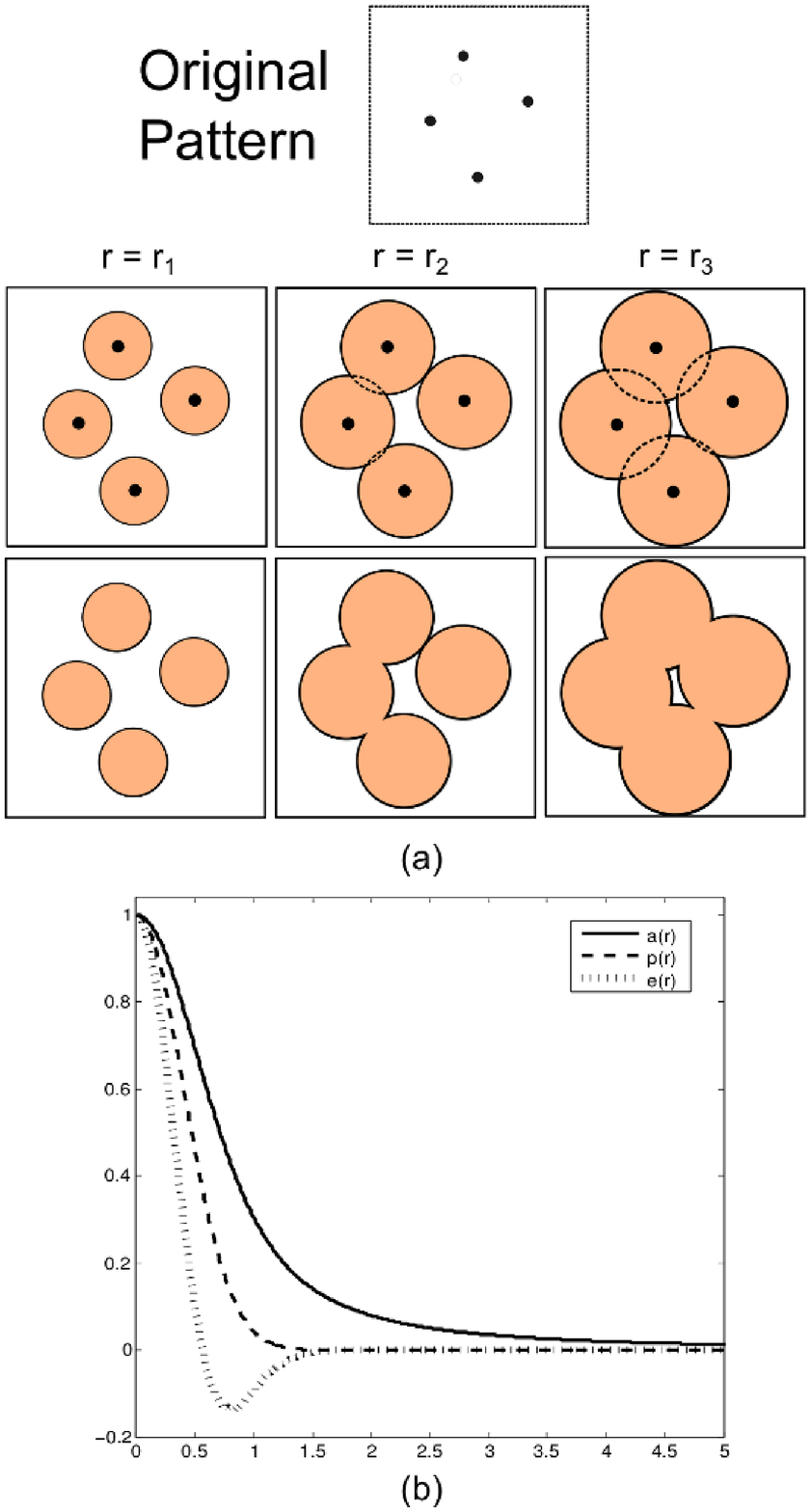}
\caption{\label{explain} (a) The Minkowski functionals are calculated by imposing discs on the point pattern. This new secondary structure can be characterized using topological measures, which vary for different radii (b) The three reduced Minkowski functionals for a 2-D Poisson (random) process. These functionals are unitless due to the normalization by the same measure one would expect for a set of non-overlapping discs}
\end{figure}

The first step in 2-D Minkowski functional analysis \cite{Mink,cosmo} is to turn a point pattern into a ''secondary pattern" by centering a disc of radius $r$ at the center of every point (see figure \ref{explain}). \footnote[1] {In this paper, we will deal only with two dimensional patterns, but our procedure is easily generalizable to patterns of any dimension.}

If the radius is large enough, some of these discs will overlap. By combining the overlapping discs, a pattern of differently shaped objects is formed. The total \emph{area}, $A$, of this collection of objects is then just the total area of the discs excluding any overlapping area. This is the first Minkowski measure. The second Minkowski measure, the total \emph{perimeter}, $P$, of the pattern is the perimeter of all of the shapes, which is again reduced from the perimeter of the individual discs because of overlaps. The \emph{Euler number}, $\chi$, is the final Minkowski measure, defined as the total number of distinct shapes or \emph{components} in the window (created by the overlapping discs) minus the number of holes.

By calculating each of these measures first for small radii, where the discs do not overlap, and growing the radius after each calculation until the entire pattern window is covered, the three Minkowski functionals ${A(r),P(r),\chi(r)}$ are approximated. Because at each radius, the Minkowski measures depend on the locations of all of the points simultaneously, these functionals include information about every type of spatial structure present in the pattern, completely characterizing it (a consequence of Hadwiger's Theorem from integral geometry, see \cite{Mecke2000}). This feature makes the Minkowski functionals a more complete measure of the underlying point interactions, including information from all possible groupings of points.

When comparing patterns, one actually uses the \emph{reduced} Minkowski functionals, namely the Minkowski functionals for the pattern divided by what is expected for a set of non-overlapping discs. These are given by
\begin{align}
  a(r) &= \frac{A(r)}{\pi N r^2}\\
  p(r) &= \frac{P(r)}{2\pi N r}\\
  e(r) &= \frac{\chi(r)}{N}
\end{align}
The functionals for a Poisson process are shown in figure \ref{explain}.b. The analysis in this paper relies exclusively on these reduced functionals, so we will not differentiate between the two. 

\FloatBarrier
\subsection{Sorting the patterns}
Our aim is to automatically sort patterns by performing FPCA on their approximated Minkowski functionals, clustering the patterns with their individual scores on the principal components. We will do the same with the pairwise correllation function so that we can directly compare our method with that of \cite{FPCA}. For each pattern set, we will use enough principal components to account for 95\% of the variation. For the Minkowski functionals, we will calculate the principal component scores individually for the \emph{area}, \emph{perimeter},and \emph{Euler number} and then concatenate the scores into a larger vector. Then, we will use these scores as coordinates, applying two different clustering algorithms: 
\begin{itemize}
\item \emph{Ward's method} \cite{ward}: An agglomerative technique which seeks to minimize the total intercluster variance of the distances between objects. We chose this method because it is well known to the pattern analysis community, and allows us to directly compare our method with that of Illian et al \cite{FPCA}.
\item \emph{Fast Weighted Modularity} \cite{modularity1,modularity2}: To implement this routine, we first calculate the pair-wise Euclidean distance between all patterns, $D_{ij}$, and transform our pattern set into a weighted graph with edge-weights
\begin{equation}
W_{ij} = max(D_{ij}) - D_{ij}.
\end{equation}
Then, this algorithm aims to maximize ''modularity" of this weighted network, by dividing the set into groups where the average edge weight between members of the same group is higher than between members of different groups. We chose this method because it is a global clustering routine with large popularity in the cluster analysis literature, and because the software implementation is able to work with very large data sets (millions of objects). 
\end{itemize}
By utilizing both cluster analysis algorithms, we can verify whatever results we obtain, and more completely demonstrate the efficacy of our sorting method using the Minkowski functionals.
\section{Testing our sorting method}
\FloatBarrier
We now apply this procedure to three different data sets. These sets of patterns highlight three possible situations in which one would sort patterns: 1) Comparing different systems, 2) Varying a parameter in an experiment, and 3) comparing different components of a bi-disperse system. (see Supplemental Material below for specifics regarding software and computational methods used). Each set is comprised of two groups which have an a priori cluster structure. We then apply the sorting technique, which allows us to calculate the percentage that is misclassified ($\%MC$) by looking at the fraction of patterns that are assigned to a group that is dominated by a different pattern type. 
\subsection{Data Set 1: Two Strauss Processes}
\FloatBarrier

\begin{figure}[ht]
\includegraphics[width=0.48\textwidth]{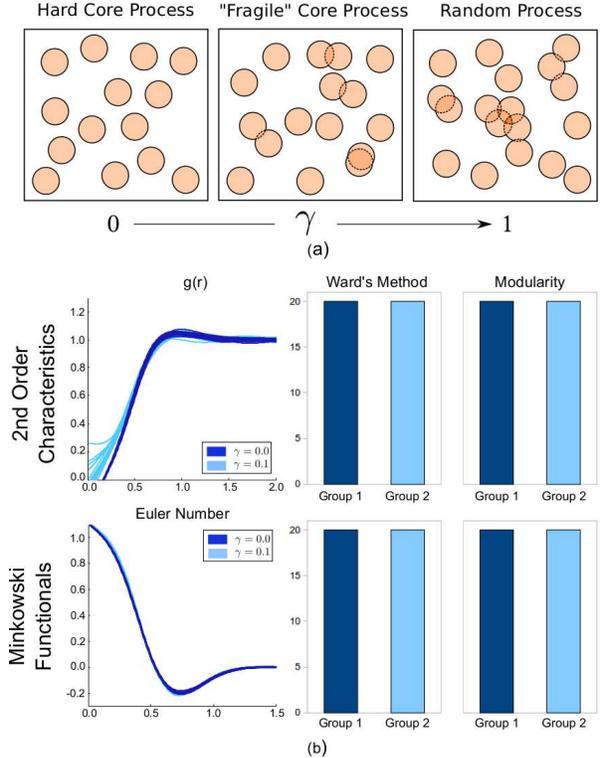}
\caption{\label{strauss} (a) The regularity of a Strauss process is completely determined by $\gamma$, the interaction parameter (b) The left most panes display the $g(r)$ and $\chi(r)$ for the 40 simulated Strauss processes. On the right are the results of using FPCA scores to divide the pattern set into two groups. As can be seen, both $g(r)$ and the Minkowski functionals can perfectly separate the set into two groups corresponding to different values of $\gamma$.}
\end{figure}

The \emph{Strauss process} \cite{strauss} is a germ-grain pattern simulation model specified by two parameters, a radius $r \in \mathbb{R}^{+}$ and an interaction parameter $\gamma \in [0,1]$. The interaction parameter determines if grains of radius $r$ will be allowed to overlap during the formation of the pattern (see figure \ref{strauss}.a).

If $\gamma$ is small, there is a strong repulsion between grains, where $\gamma = 0$ yield a hard-core process. If $\gamma$ is close to $1$, the repulsion is weak, where $\gamma = 1$ yields a completely random process. 
In \cite{FPCA}, it was reported that even for comparison of pattern sets with similarly strong repulsion ($\gamma = 0.0$ and $\gamma = 0.1$) the pairwise correllation function was able to effectively distinguish different Strauss processes. We here repeat this test with 20 patterns each, fixing the number of points at $N = 1000$, and letting $r = 0.025$. We also fix the number intensity to be $\lambda = (2r)^{-2}$, which forces interaction between the points. 

As can be seen in figure \ref{strauss}.b, both $g(r)$ and the Minkowski functionals are able to distinguish the two Strauss processes, separating the pattern set into two homogenous groups. This is to be expected for $g(r)$, as 2nd-order interactions dominate the process, and is consistent with the findings of \cite{FPCA}.

\subsection{Data Set 2: Baddelley-Silverman vs. Random}
The \emph{Baddelley-Silverman process} \cite{baddelley} is built by partitioning a domain into a grid and moving from box to box, distributing $N$ points in each box uniformly. $N$ itself is a random number, taking on the values $0$,$1$, and $10$ with probabilities $1/10$, $8/9$, and $1/90$, respectively. This causes the process to be rather regular, but with some strong clustering occuring every now and then (see figure 6).
\begin{figure}
\includegraphics[width=0.47\textwidth]{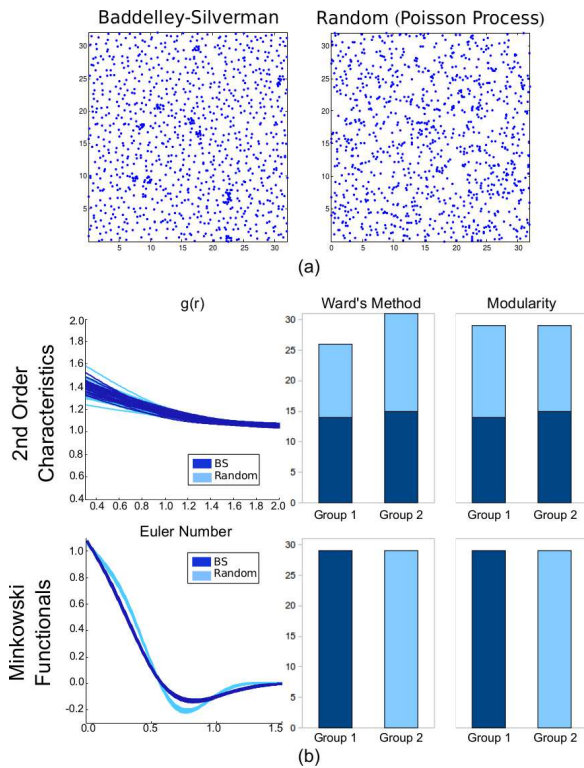}
\caption{\label{baddelley} (a) A Baddelley-Silverman process side-by-side with a Poisson process. Despite the visible differences, the pairwise correlation functions are identical (b) The left most panes display the $g(r)$ and $\chi(r)$ for the 58 patterns simulated. On the right are the results of using FPCA scores to divide the pattern set into two groups. As can be seen, FPCA sorting with $g(r)$ creates two perfectly heterogenous groups, while FPCA sorting with the Minkowski functionals groups the patterns correctly}
\end{figure}

Since $E[N] =$ Var$[N] = 1$, it can be shown that the Baddelley-Silverman process shares all of the same 2nd-order characteristics as a random (CSR) process. In ref. \cite{Mink}, the Minkowski functionals were shown to be  able to distinguish these two processes. Therefore, we expect to see proper sorting when using $a(r)$, $p(r)$, and $e(r)$, and failure using $g(r)$.

We simulated 29 BS processes and 29 Poisson processes, fixing the point number N = 1024. Using both the pairwise correlation function and the Minkowski functionals, and sorted the pattern sets to into two groups. As can be seen in figure \ref{baddelley}.b, the pairwise correllation function fails to sort the patterns correctly, creating two heterogenous groups (\%MC $> 40$). However, the Minkowski functionals successfully divide the pattern set into two homogenous groups.

\subsection{Data Set 3: Bi-disperse patterns of inter-cellular proteins}
\begin{figure}
\includegraphics[width=0.47\textwidth]{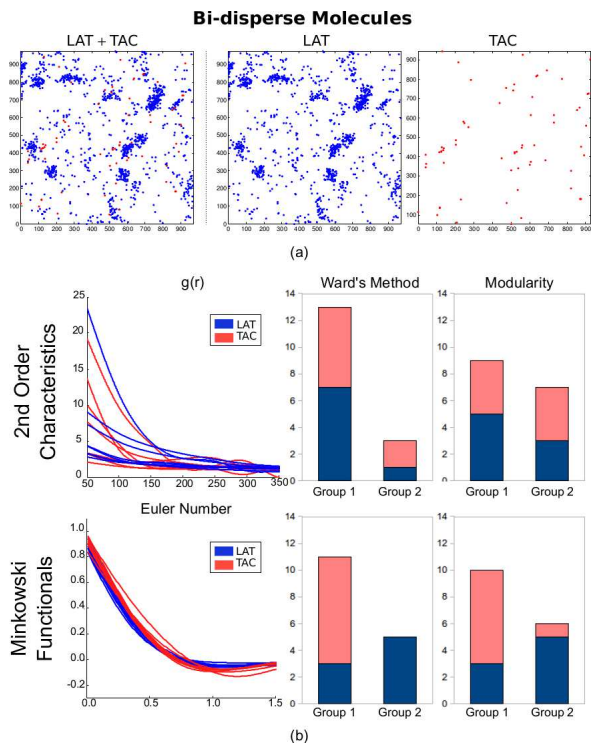}
\caption{\label{proteins} (a) The two proteins are both dispersed in the cell membrane, but can be visualized separately (b) The left most panes display the $g(r)$ and $\chi(r)$ for the 16 molecular patterns. On the right are the results of using FPCA scores to divide the pattern set into two groups. As can be seen, using Minkowski functionals with FPCA improves the differentiation of the two sets}
\end{figure}

For an application to an experimental data set, we look at super-resolution images of two proteins residing at the membrane of immune cells (see figure \ref{proteins}.a). 

One protein under study is LAT, short for ''Linker for Activation of T-cells", a naturally occuring protein crucially involved in the reactions that regulate T-Cell antigen-dependent activation, a critical event in the adaptive immune response. LAT proteins have been seen to form clusters on the membrane with potentially complicated hierarchies \cite{Eilon}. However, the membrane of the cell can have a 1st-order effect on the molecular patterning of membrane proteins. It has been found that the location of other membrane protein clusters often correlates with how close the membrane is to the surface, and anti-correlates with regions of high membrane fluctuations \cite{arpita}.

Another protein, TAC (the alpha chain of the IL-2 receptor), can also be localized and differentiated from LAT by tagging with a different fluorescent molecule and using two different lasers with different wavelengths. TAC is a membrane protein that does not form clusters, instead distributing uniformly in regions where protein-membrane interactions have not excluded proteins. This means that TAC can serve as a membrane marker when studying the clustering of other proteins. Since LAT and TAC are part of separate signalling pathways, they also do not interact biochemically \cite{lattac}. Therefore, upon sorting, we should get two homogenous groups representing the two different molecules.

Applying FPCA on the approximated pairwise correlation functions of these data sets again yields strongly heterogeneous groups  (\%MC $\approx$ 50\%). This is visible in the pair-correlation functions (figure \ref{proteins}.b), where the individual patterns exhibit large variability. In contrast, because the Minkowski functionals consider more than just 2nd-order interactions, the $Euler$ $number$ is able to visibly distinguish the molecular patterns, and the pattern sorting is improved (\%MC $\approx$ 25\%). 
\begin{figure}
\includegraphics[width=0.5\textwidth]{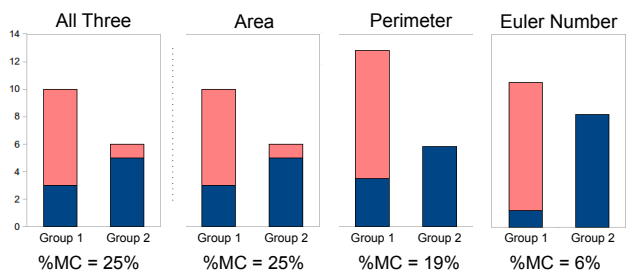}
\caption{\label{mink_results} Considering the individual functionals (sorting using weighted modularity), the Euler number outperforms the other two, only misclassifying one pattern}
\end{figure}
Further success is achieved if we look at how FPCA sorting with each functional performs on its own (see figure \ref{mink_results}). When only using the $area$, the sorting is identical to the sorting based on all three Minkowski functionals. However, sorting the LAT/TAC protein pattern set improves when just using the $perimeter$, and we achieve almost perfect classification when using the $Euler$ $number$, only misclassifying one pattern. This is not surprising, since the $area$ and $perimeter$ are constrained to be smooth, positive, and monotonically decreasing, and thus cannot vary as much while the $Euler$ $number$ can vary more wildly.

\section{Conclusions and Discussion}
In this work, we have introduced the procedure to automatically sort point pattern sets using the approximated Minkowski functionals and Functional Principal Component Analysis (FPCA). Using Strauss processes with strong repulsion, we have shown that this method can accurately sort point pattern sets drawn from very similar processes. Further, this method also distinguishes Baddeley-Silverman processes from Poisson processes, a task which the pairwise correlation function perfectly fails to accomplish. We then found that when looking at experimental point patterns representing proteins, FPCA sorting using the Minkowski functionals outperformed FPCA sorting with the pairwise correlation function. This sufficiently demonstrates that the Minkowski functionals can successfully quantify the differences between pattern sets showing complex behavior.

We also found that FPCA sorting using only the \emph{Euler number} strongly outperforms the other two. While mathematically the three functionals do completely classify a pattern, the $area$ and \emph{perimeter} may only be slightly different for different spatial processes. This means that error introduced when approximating the functionals numerically may blur these differences, resulting in improper sorting. Since the $Euler$ $number$ is allowed to vary more dramatically as the discs combine and holes form, it can visually distinguish very similar pattern sets, and therefore leads to better sorting.

Though we have presented this technique as a way to sort patterns into distinct sets for further analysis, the sorting itself can serve as an analysis tool. We are currently working to apply this tool to examine how the presence of different chemical cues effect the clustering of LAT proteins, as well as how T-cell activation perturbs the patterns. Because of the Minkowski functional's ability to robustly characterize a pattern, we can treat the membership of a pattern in a particular group as a sign of similarity between it and it is co-members. We can therefore look at group statistics to determine what experimental variables change the molecular patterns, and to what degree, allowing for systematic large-scale investigations of the membrane proteins and their response to different stimuli.

\section{Acknowledgments}
The authors would like to thank Detlef Lohse for suggesting Minkowski Functional analysis, and Can Guven and Mark Herrera for their helpful advice during this project. This research was supported in part (LES) by the Intramural Research Programs of the National Cancer Institute (The Center for Cancer Research) and by NIH grant GM085574 (WL).

\section{Supplemental Material}

\subsection{Software}
To approximate the 2-D Minkowski functionals of our patterns, we relied exclusively on the software described in \cite{Mink}, which was available online at \url{http://www.mathe.tu-freiberg.de/inst/stoch/Stoyan/morph2D/}. \footnote{At the time of this paper's submission, this website was down; we are in the process of notifying the appropriate people about this issue}. This program takes as input $rmin$, $dr$, and $rmax$. Since our interest is in automation we used the same values for all of our patterns ($r_{min} = dr = .01, r_{max} = 100$).

For both smoothing and applying the Functional Principal Component Analysis, we used the Functional Data Analysis MATLAB packages that are available online at \url{www.functionaldata.org}, and we relied on their description in \cite{ramsay2} for implementation. Mimicking the procedure of \cite{FPCA}, we first smoothed our functionals using cubic b-splines.

To cluster using Ward's method, we first utilized MATLAB's implementation in their ''linkage" function. To implement modularity maximization, we used the weighted version of the Fast Modularity algorithm which can be found online at \url{http://cs.unm.edu/~aaron/research/fastmodularity.htm}. The specifics of the algorithm are the same as in \cite{modularity1}, but maximizes the weighted definition of modularity (for a description of this alteration, see \cite{modularity2}). 

Other home made programs were written to compute the second order functionals, simulate point patterns, and implement various portions of the project (either in MATLAB or C). Those interested in discussing these programs should contact the authors.
\subsection{Intensity Scaling}
As reported in \cite{Mecke2000}, the Minkowski functionals are homogenous with regard to domain scaling. To be specific, for any parameter $\lambda > 0$ and domain $\Omega \subset \mathbb{R}^d$, the $n$-th Minkowski functional $M_n (|\Omega|)$ satisfies the relation
\begin{equation}
M_n (|\lambda \Omega|) = \lambda^{d-n} M_n(|\Omega|)
\end{equation}
This means two patterns with different overall number intensity will have different Minkowski functionals even if they are the same type of pattern. To address this in our pattern comparison, all patterns were scaled to unit intensity before their Minkowski functionals were approximated. 
\subsection{Approximating $g(r)$}
To approximate the pairwise correlation function $g(r)$, we used the estimator
\begin{equation}
\hat{g}(r) = N^{-1}\lambda^{-1}\displaystyle\sum\limits_{i=1}^N w_i^{-1}(r,dr) \displaystyle\sum\limits_{i\not= j} I(|\vec{r}_i - \vec{r}_j| < r).
\end{equation}
Here, $I(x)$ is the indicator function and $\lambda$ is the number intensity. The weight $w_i$ is the portion of the area of the disc centered on $\vec{r}_i$ with inner radius $r$ and outer radius $r + dr$ that is contained in the pattern window. We found that this method achieved better results than that of \cite{FPCA}, where $g(r)$ is approximated by exploiting it's relation to the derivative of Ripley's K-function.
\subsection{Pattern Simulation}
Binomial processes were simulated using MATLAB's built in random number generator and scaling the results. MATLAB code to simulate Strauss processes can be found in \cite{strauss_sim}, and Baddelley-Silverman processes were simulated using homemade software based on the procedure described in \cite{baddelley}. 

\bibliographystyle{ieeetr}
\bibliography{myrefs}

\begin{thebibliography}{10}

\bibitem{ecology}
R.~Law, J.~Illian, D.~F. R.~P. Burslem, G.~Gratzer, C.~V.~S. Gunatilleke, and
  I.~A. U.~N. Gunatilleke, ``{Ecological information from spatial patterns of
  plants: insights from point process theory},'' {\em Journal of Ecology},
  vol.~97, no.~4, pp.~616--628, 2009.

\bibitem{wiegand}
T.~Wiegand and K.~Moloney, ``Rings, circles, and null-models for point pattern
  analysis in ecology,'' {\em Oikos}, vol.~104, no.~2, pp.~209--229, 2004.

\bibitem{crime}
G.~O. Mohler, M.~B. Short, P.~J. Brantingham, F.~P. Schoenberg, and G.~E. Tita,
  ``{Self-Exciting Point Process Modeling of Crime},'' {\em Journal of the
  American Statistical Association}, vol.~106, pp.~100--108, Mar. 2011.

\bibitem{epid}
A.~C. Gatrell, T.~C. Bailey, P.~J. Diggle, and B.~S. Rowlingson, ``{Spatial
  Point Pattern Analysis and Its Application in Geographical Epidemiology},''
  {\em Transactions of the Institute of British Geographers}, vol.~21, no.~1,
  pp.~256--274, 1996.

\bibitem{econ}
R.~Bivand, ``A review of spatial statistical techniques for location studies,''
  in {\em Norwegian School of Economics and Business Administration},
  pp.~1998--98, 1998.

\bibitem{seis}
Y.~Ogata, ``Seismicity analysis through point-process modeling: A review,''
  {\em Pure and Applied Geophysics}, vol.~155, pp.~471--507, 1999.

\bibitem{material}
S.~J.~L. {Billinge}, ``{The atomic pair distribution function: past and
  present},'' {\em Zeitschrift fur Kristallographie}, vol.~219, pp.~117--121,
  Mar. 2004.

\bibitem{astro}
W.~Kendall, ``Statistics of the galaxy distribution,'' {\em Journal of the
  American Statistical Association}, vol.~98, pp.~248--248, January 2003.

\bibitem{PALM}
E.~Betzig, G.~Patterson, R.~Sougrat, O.~Lindwasser, S.~Olenych, J.~Bonifacino,
  M.~Davidson, J.~Lippincott-Schwartz, and H.~Hess, ``Imaging intracellular
  fluorescent proteins at nanometer resolution.,'' {\em Science (New York)},
  vol.~313, no.~5, pp.~1642--5, 2006.

\bibitem{coloc}
E.~Lachmanovich, D.~E. Shvartsman, Y.~Malka, C.~Botvin, Y.~I. Henis, and A.~M.
  Weiss, ``Co-localization analysis of complex formation among membrane
  proteins by computerized fluorescence microscopy: application to
  immunofluorescence co-patching studies,'' {\em Journal of microscopy},
  vol.~212, pp.~122--131, Nov. 2003.

\bibitem{Eilon}
E.~Sherman, V.~Barr, S.~Manley, G.~Patterson, L.~Balagopalan, I.~Akpan, C.~K.
  Regan, R.~K. Merrill, C.~L. Sommers, J.~Lippincott-Schwartz, and L.~E.
  Samelson, ``Functional nanoscale organization of signaling molecules
  downstream of the t cell antigen receptor,'' {\em Immunity}, Nov. 2011.

\bibitem{Stat}
J.~Illian, A.~Penttinen, H.~Stoyan, and D.~Stoyan, {\em Statistical analysis
  and modeling of Spatial Point Patterns}.
\newblock Statistics in Practice, John Wiley and Sons, 2008.

\bibitem{FPCA}
J.~Illian, E.~Benson, J.~Crawford, and H.~Staines, ``Principal component
  analysis for spatial point processes — assessing the appropriateness of the
  approach in an ecological context,'' in {\em Case Studies in Spatial Point
  Process Modeling} (A.~Baddeley, P.~Gregori, J.~Mateu, R.~Stoica, D.~Stoyan,
  P.~Bickel, P.~Diggle, S.~Fienberg, U.~Gather, I.~Olkin, and S.~Zeger, eds.),
  vol.~185 of {\em Lecture Notes in Statistics}, pp.~135--150, Springer New
  York, 2006.

\bibitem{ramsay1}
J.~O. Ramsay and B.~W. Silverman, {\em {Functional Data Analysis}}.
\newblock Springer Series in Statistics, Springer, 2nd~ed., June 2005.

\bibitem{ramsay2}
J.~O. Ramsay, G.~Hooker, and S.~Graves, {\em Functional Data Analysis with R
  and MATLAB}.
\newblock Springer Publishing Company, Incorporated, 1st~ed., 2009.

\bibitem{diggle}
J.~P. Diggle, {\em {Statistical Analysis of Spatial Point Patterns}}.
\newblock Academic Press, New York, 1983.

\bibitem{stoyan}
A.~Tscheschel and D.~Stoyan, ``Statistical reconstruction of random point
  patterns,'' {\em Comput. Stat. Data Anal.}, vol.~51, no.~2, pp.~859--871,
  2003.

\bibitem{baddelley}
A.~J. Baddeley and B.~W. Silverman, ``A cautionary example on the use of
  second-order methods for analyzing point patterns,'' {\em Biometrics},
  vol.~40, no.~4, pp.~pp. 1089--1093, 1984.

\bibitem{Mink}
K.~R. Mecke and D.~Stoyan, ``Morphological characterisation of point
  patterns,'' {\em Biometrical Journal}, vol.~47, no.~4, pp.~473--488, 2005.

\bibitem{cosmo}
J.~Schmalzing, M.~Kerscher, and T.~Buchert, ``{Minkowski Functionals in
  Cosmology},'' in {\em Dark Matter in the Universe} (S.~Bonometto, J.~Primack,
  and A.~Provenzale, eds.), p.~281, 1996.

\bibitem{aa}
M.~Kerscher and A.~Tikhonov, ``Morphology of the local volume,'' {\em A\&A},
  vol.~509, p.~A57, 2010.

\bibitem{alum}
S.~Gallier, ``A stochastic pocket model for aluminum agglomeration in solid
  propellants,'' {\em Propellants, Explosives, Pyrotechnics}, vol.~34, no.~2,
  pp.~97--105, 2009.

\bibitem{detlef}
E.~Calzavarini, M.~Kerscher, D.~Lohse, and F.~Toschi, ``Dimensionality and
  morphology of particle and bubble clusters in turbulent flow,'' {\em Journal
  of Fluid Mechanics}, vol.~607, pp.~13--24, 2008.

\bibitem{Mecke2000}
K.~R. Mecke, ``{Additivity, convexity, and beyond: applications of Minkowski
  functionals in statistical physics},'' {\em Lecture Notes in Physics},
  vol.~554, pp.~111--184, 2000.

\bibitem{ward}
J.~Ward, ``Hierarchical grouping to optimize an objective function.,'' {\em
  Journal of the American Statistical Association}, vol.~58, pp.~236--244,
  1963.

\bibitem{modularity1}
A.~Clauset, M.~E.~J. Newman, and C.~Moore, ``Finding community structure in
  very large networks,'' {\em Phys. Rev. E}, vol.~70, p.~066111, Dec 2004.

\bibitem{modularity2}
M.~E.~J. Newman, ``Analysis of weighted networks,'' {\em Phys. Rev. E},
  vol.~70, p.~056131, Nov 2004.

\bibitem{strauss}
D.~J. Strauss, ``A model for clustering,'' {\em Biometrika}, vol.~62, no.~2,
  pp.~467--475, 1975.

\bibitem{arpita}
K.~Lam~Hui, C.~Wang, B.~Grooman, J.~Wayt, and A.~Upadhyaya, ``{Membrane
  Dynamics Correlate with Formation of Signaling Clusters during Cell
  Spreading},'' {\em Biophys J}, vol.~102, pp.~1524--1533, Apr. 2012.

\bibitem{lattac}
R.~Robb, W.~Greene, and C.~Rusk, ``Low and high affinity cellular receptors for
  interleukin 2. implications for the level of tac antigen.,'' {\em J Exp Med},
  vol.~160, no.~4, pp.~1126--46, 1984.

\bibitem{strauss_sim}
W.~Martinez, {\em Computational statistics handbook with MATLAB}.
\newblock Chapman \& Hall/CRC, 2001.

\end{thebibliography}

\end{document}